\documentclass[12pt]{article}
\usepackage[dvips]{graphicx}
\usepackage{amsfonts}
\usepackage{amsmath}

\title{A simple method for the evaluation of the information content and complexity in atoms.
\\A proposal for scalability.}

\author{C.P. Panos$^{a}$, N.S. Nikolaidis$^{b}$, K.Ch. Chatzisavvas$^{a}$, C.C.
Tsouros$^{c}$
\\
 {\small\it  $^{a}$Physics Department, Aristotle University of Thessaloniki,}\\
 {\small\it  54124 Thessaloniki, Greece.}
\\
 {\small\it  $^{b}$Department of Automation, Faculty of Applied Technology,}\\
 {\small\it  Alexander Technological Educational Institute (ATEI) of Thessaloniki,}\\
 {\small\it  57400 Thessaloniki, Greece.}
\\
 {\small\it  $^{c}$School of Mathematics, Physics and Computational Sciences,}\\
 {\small\it  Faculty of Engineering, Aristotle University of Thessaloniki,}\\
 {\small\it  54124 Thessaloniki, Greece.}
}
\begin{document}
\date{November, 2008}
\maketitle

\begin{abstract}
We present a very simple method for the calculation of Shannon,
Fisher, Onicescu and Tsallis entropies in atoms, as well as SDL
and LMC complexity measures, as functions of the atomic number
$Z$. Fractional occupation probabilities of electrons in atomic
orbitals are employed, instead of the more complicated continuous
electron probability densities in position and momentum spaces,
used so far in the literature. Our main conclusions are compatible
with the results of more sophisticated approaches and correlate
fairly with experimental data. We obtain for the Tsallis entropic
index the value $q=1.031$, which shows that atoms are very close
to extensivity. A practical way towards scalability of the
quantification of complexity for systems with more components than
the atom is indicated. We also discuss the issue if the complexity
of the electronic structure of atoms increases with $Z$. A Pair
($\alpha, \beta$) of Order-Disorder Indices (PODI), which can be
introduced for any quantum many-body system, is evaluated in atoms
($\alpha=0.085,\beta=1.015$). We conclude that "atoms are ordered
systems, which do not grow in complexity as $Z$ increases".
\end{abstract}

%\begin{keyword}
%\PACS 89.70.Cf
%\end{keyword}

\section{Introduction}\label{sec:intro}

Information-theoretic methods have been used extensively to study
various systems in communications \cite{ShannonArticle}, physics
\cite{BBM}, chemistry \cite{SearsPHD}, \cite{PaperKP}, biology
\cite{AdamiPLR}, e.t.c. Applications in quantum systems, e.g.
atoms (\cite{GadreSearsEtAlPaper}, \cite{SenEdBook} and references
therein), have been rewarding and applied with considerable
success. Also Shannon information entropy has been connected with
experiment, e.g. the ionization potential and dipole
polarizability \cite{PaperSPCM}. Related research in atoms employs
Hartree-Fock wavefunctions, obtained numerically, which can be
used as an input for the calculation of Shannon information
entropies in position- and momentum- spaces according to the
definitions:

\[
    S_{r}=-\int{\rho (\textbf{r}) \, \ln \rho(\textbf{r}) \, d \textbf{r}}
    \quad \mbox{and} \quad
    S_{k}=-\int{n(\textbf{k}) \, \ln n(\textbf{k}) \, d \textbf{k}}
\]

\noindent where $\rho(\textbf{r})$ and $ n(\textbf{k})$ stand for
the electron probability densities in position- and momentum-
spaces respectively. Entropic Uncertainty Relations (EUR)
\cite{BBM} hold of the form

\[
    S_{r}+S_{k}\geq n(1+\ln \pi)
\]

\noindent($n=3$ for a 3-dimensional system). EUR are considered as
an improvement compared with the usual Heisenberg uncertainty
relations, in the following sense: First, one can derive
Heisenberg's relations from EUR, whereas the inverse is not true
and, second, the right-hand side of EUR does not depend on the
state of the system, whereas Heisenberg's do depend. Another
important issue is that the entropic sum $S = S_{r} + S_{k}$ is
scale-invariant i.e. it does not depend on the units chosen to
measure position $\textbf{r}$ and momentum $\textbf{k}$. Thus, $S$
is an important dimensionless quantity that represents the
information content of a quantum system in bits, if the base of
the logarithm in the definition of $S_{r}$ and $S_{k}$ is 2, nats,
if the base is e, and Hartleys, for base 10. Here, we use the
natural logarithm.

An interesting universal property is: $ S=a+b \ln N$, where $N$ is
the number of particles of a quantum system. So far, it has been
shown empirically to hold approximately for various quantum
systems \cite{GadreSearsEtAlPaper}, \cite{PaperMMP},
\cite{PaperMP2}, such as atoms, nuclei, atomic clusters and bosons
in a trap. These are systems with various numbers of particles
$N$, ranging from a few to millions, and have various sizes, with
constituent particles obeying different interactions and different
statistics such as fermions and bosons. The parameters $a$, $b$
depend on the system under consideration. $S$ is connected with
the kinetic energy \cite{PaperMP}. Another important finding is
the discovery that the application of external electric and
magnetic fields to an atom influences the information content of
electrons \cite{FerezDehesa.Paper}. Complexity in atoms has been
quantified, starting in \cite{PaperCMP} and extending in
\cite{PaperPCMK}. Thus, in \cite{PaperCMP}, for the first time,
complexity was evaluated as an index characterizing a quantum
system, specifically an atom. The SDL (Shiner, Davison, Landsberg)
\cite{SDL.paper} and LMC (Lopez-Ruiz, Mancini, Calbet)
\cite{LMC.paper} \emph{statistical} measures of complexity were
employed, defined in such a way as to overcome the fact that
sometimes entropy is not a proper measure of disorder and
complexity. They are called \emph{statistical}, because they are
based on information entropy, calculated from probability
densities. The question arises, whether one can predict that a
system has the ability to organize itself, i.e. increase its
complexity, as the number of particles $N$ increases or another
parameter changes, or it needs some external factor or agent
towards self-organization (see discussion in \cite{PaperPCMK}).

A difficult scientific and philosophical issue is to analyze a
system in terms of interacting components (reductionism) or
inversely, to derive the properties of the initial system from
those of the constituents. The aim is to investigate
quantitatively whether reductionism is correct or if, by moving to
higher levels in the hierarchy of systems (bottom-up), new laws
appear, which cannot be explained in terms of components lying
lower in the hierarchy. For example, the detailed calculation of
electronic properties of molecules is cumbersome or intractable. A
real progress, in that case, became feasible in the past, when an
effective hypothesis was made, i.e. the chemical bonding,
resulting in an immense and practical understanding of molecules,
with infinite applications, without the need to know or calculate
all the details of the underlying electronic structure. An example
of quantification of the order of the chemical bonding using
Shannon information is \cite{PaperKP}. Specifically, an
information-theoretic comparison of Fermi and Coulomb electron
pairs was carried out, employing a simplified probabilistic
procedure.

Researchers could desire to understand the following hierarchy:\\
\emph{Nucleons} $\rightarrow$ \emph{Nuclei} $\rightarrow$
\emph{Atoms} $\rightarrow$ \emph{Molecules} $\rightarrow$
\emph{Larger Structures (e.g. DNA)}. \\ In each level of
description, information is processed in different ways, with
various degrees of freedom and constants being relevant. One could
not be sure that he/she will ever be capable of crossing the
border from one level to the next, even in principle.
Pragmatically speaking, detailed numerical calculations are
difficult for the above transition and probably useless. One could
hope to find simple methods to do this, without compromising the
important characteristics of systems. In this paper, we propose
such a simple method for atoms, reproducing the main results
obtained from more sophisticated and accurate models, leading to
the same qualitative conclusions for self-organized complexity.
The next molecular level might be reached if, for example, the
probabilities of electrons to occupy basins in various molecules
become available and used in a simplified probabilistic treatment
of information and complexity, instead of more complicated
calculations. At the moment, the electronic structure of molecules
is obtained \emph{ab initio} or from approximate calculations,
which are quite involved and mostly carried out for a specific
molecule. A simplified systematic approach is called for. As we
proceed to higher levels in the hierarchy of systems, we should
check whether the basic features, related to information and
complexity, are preserved by a simplified approach trying to
achieve a controlled scalability.

An approach in a spirit similar to our present work is the
treatment of information and complexity by Bonchev in Ch. 4 of
\cite{BonchevBook}. In fact, our starting step to employ
fractional occupation probabilities of atomic orbitals in order to
evaluate the Shannon entropy in atoms is analogous with
\cite{BoncevKamenskaArticle}, but, here, we elaborate further, by
extending to SDL and LMC measures of complexity and Tsallis
entropy. Another point of view is a survey of several molecular
indices, based on evaluating graph complexity \cite{BonchevBook}.
We also mention that calculations of Shannon entropy for molecules
have already been carried out using molecular wavefunctions
\cite{MinhhuyPaper}, \cite{MinhhuyPaper2}. A very recent
calculation of information and complexity measures for H$_{2}+$
may serve as a promising starting point \cite{KalidasH2}. The aim
of \cite{KalidasH2} is to clarify the nature of molecular bonding,
employing information-theoretic and complexity concepts. Finally,
one should not omit the comprehensive information-theoretic
treatment of molecular systems in \cite{NalewajskiBook} and
references therein.

The outline of the paper is the following: In Section
\ref{sec:Definitions}, we give various definitions of Information
and Complexity measures, while in Section \ref{sec:Methodology} we
present our method and numerical results. Specifically, subsection
\ref{subsec:SDLcoeff} contains a new definition of a Pair of
Order-Disorder Indices (PODI), characterizing quantum many-body
systems \cite{ChatzPanosPODI}. Section \ref{sec:Discussion}
contains a discussion and finally, in Section
\ref{sec:Conclusions}, we display our main conclusions.

\section{Definitions of Information and Complexity Measures}
\label{sec:Definitions}

Let us consider a discrete probability distribution $ {p_{i}},
i=1,2,\ldots ,\nu $. The corresponding Shannon information entropy
\cite{ShannonArticle} is defined as:

\begin{equation}
    S = -\sum_{i} p_{i} \ln p_{i}
\label{eq:ShannonDefinitionDiscrete}
\end{equation}

\noindent where $\sum_{i} p_{i} = 1$. In the case of atoms, $
p_{i} $ might be the occupation probability of a specific electron
orbital. The maximum and minimum values of $S$, in this case, will
be:

\begin{equation}
    S_{max} = \ln \nu,\quad S_{min}=0
\label{eq:ShannonDefMaxDiscrete}
\end{equation}

\noindent where $\nu$ is the number of occupied orbitals, while
$S_{min}=0$ holds only if one of the $p_{i}$'s equals 1 and all
the rest equal 0. We note that for continuous density
distributions e.g. in atoms or other quantum many-body systems
i.e. nuclei, atomic clusters and bosons in a trap, $S_{min}$ and
$S_{max}$ obey the much more complicated inequalities of Gadre
\cite{GadreSearsEtAlPaper}, \cite{SenEdBook} instead of
(\ref{eq:ShannonDefMaxDiscrete}). These inequalities were verified
in \cite{PaperMMP}. $S$ is a global measure of information in the
sense that, by changing the order of the values $ p_{i},
i=1,2,\ldots ,\nu $, there is no effect on the value of the sum in
relation (\ref{eq:ShannonDefinitionDiscrete}). On the contrary,
Fisher's definition of information \cite{FisherArticle},
\cite{FriedenBook}

\begin{equation}
    I=\int \frac{\left( \frac{\rho(\scriptsize{\textbf{r}})}{d \scriptsize{\textbf{r}}} \right)^2}
    {\rho(\textbf{r})}\,d\textbf{r}
\label{eq:FisherDefinitionContinuous}
\end{equation}

\noindent is called a local measure of information, because it
contains the derivative of the continuous density distribution
$\rho(\textbf{r})$. The same argument can be extended for a
discrete distribution, where the corresponding Fisher information
is:

\begin{equation}
    I=\sum_{i=1}^{\nu}{\frac{(p_{i+1}-p_{i})^{2}}{p_{i}}}
\label{eq:FisherDefinitionDiscrete}
\end{equation}

\noindent where we put $p_{\nu+1}=0$. Here, one should use a
specific sequence of the probabilities $\{p_{i}\}$ and the obvious
choice is the ordering of atomic orbitals in electron
configurations, dictated by nature. Recently in \cite{PaperCMP},
\cite{PaperPCMK}, the information entropy $ S=S_{r}+S_{k} $ for
atoms, derived probabilistically, was used to calculate
quantitatively complexity measures. First, the SDL measure
\cite{SDL.paper} was calculated as a function of $Z$, defined as:

\begin{equation}
    \Gamma_{\alpha,\beta} = \Delta ^{\alpha} \cdot \Omega^{\beta}
\label{eq:SDLdefinition}
\end{equation}

\noindent where $\Delta$ is the disorder of the system (Landsberg)
\cite{LandsbergBook}, \cite{LandsbergArticle}:

\begin{equation}
    \Delta = \frac{S}{S_{max}}
\label{eq:DeltaDef}
\end{equation}

\noindent and $\Omega$ is the order:

\begin{equation}
\Omega = 1 -\Delta \label{eq:OmegaDef}
\end{equation}

In other words, $\Delta$ and $\Omega$ are the normalized measures
of disorder and order respectively, with $0<\Delta<1$ and
$0<\Omega<1$.

The SDL measure of complexity obeys the desired property:
$\Gamma_{\alpha,\beta} = 0$ for both extreme cases of an ideal gas
in complete disorder, $S=S_{max}$ and a perfect crystal  in
complete order, $S=0$. The interesting part is between complete
order and complete disorder, which is intuitively satisfactory,
i.e.

\[
    0 < \Gamma _{\alpha,\beta} < 1
\]

The values $\alpha $, strength of disorder, and $\beta$, strength
of order, are still to be specified and play a crucial role to
observe an increasing, decreasing or convex trend for $
\Gamma_{\alpha ,\beta} $ as a function of $ \Delta $ or $ \Omega $
or the number of particles $N$ \cite{PaperPCMK}, \cite{SDL.paper}.

The particular values of $ \alpha $ and $ \beta $ in
(\ref{eq:SDLdefinition}) might be chosen by imposing the condition
that the values of the two complexity measures $ \Gamma_{\alpha,
\beta} $ and $ C $ are approximately the same. This comparison has
been suggested in \cite{PaperPCMK}, where we obtained roughly $
\alpha \simeq 0 $ and $ \beta \simeq 4 $ and has been observed
that complexity increases with $Z$ based on the trend of closed
shells. However, maybe the solution is not unique and it would be
better to find various regions of pairs $(\alpha , \beta)$ in the
$\alpha-\beta$ plane, each one characterized by an index giving a
different behavior (increasing, decreasing or convex) for
complexity, as a function of N. A more detailed search for the
proper values of $(\alpha,\beta)$ is being carried out in a paper
in preparation \cite{ChatzPanosPODI}, employing more accurate
continuous densities in position and momentum spaces applied in
atoms and other quantum many-body systems as well. This procedure
may lead to a quantitative answer, whether the system can show
self-organization without the intervention of external factors.
The effect of external electric and magnetic fields on the
information content of some excited states of hydrogen was studied
in \cite{FerezDehesa.Paper}. A first step to evaluate the effect
of the \emph{environment} is to study confined atoms
\cite{SenPaper}.

Another measure of complexity is the LMC measure \cite{LMC.paper}:

\begin{equation}
    C= S  \cdot D
\label{eq:LMCdefinition}
\end{equation}

\noindent where $S$ is the usual information entropy, while $D$ is
the disequilibrium, i.e. the distance of the specific
non-equiprobable distribution ${p_{i}}$ from a uniform
distribution $p_{i }= \frac{1}{\nu}, i=1 \ldots \nu $, where $\nu$
is the number of occupied orbitals of the atom. Thus, for the
discrete case:

\begin{equation}
    D = \sum_{i=1}^{\nu} (p_{i}-\frac{1}{\nu})^{2}
\label{eq:DisequilibriumDefinition1Discrete}
\end{equation}

The definition of the Onicescu information energy for a continuous
distribution is $ E = \int{\rho ^2(\textbf{r}) d\textbf{r}} $ and
for a discrete one is:

\begin{equation}
    E = \sum_{i=1}^{\nu} p_{i}^2
\label{eq:DisequilibriumDefinition2Discrete}
\end{equation}

In the literature, the quantity $E$ is called Onicescu's
information energy \cite{OnicescuBook},
\cite{OnicescuStefanescuBook}, although it does not have the
dimension of energy. This name has been given by analogy with
thermodynamics, because E attains a minimum for a uniform
distribution of equal probabilities --total disorder. It can be
employed as a finer measure of information content
\cite{PaperCMP}. We also note that the continuous generalization
of $D$ is $D=\int{\rho^{2}(\textbf{r})d{\textbf{r}}}$, the same
with $E$. $D$ is an experimentally measurable quantity
\cite{HymanPaper}, known also as quantum self-similarity
\cite{BorgooPaper2}.

The definition of information content given above can be extended
employing the Tsallis entropy \cite{Tsallis.paper}:

\begin{equation}
    T_{q} = \frac{1-\sum_{i} p_{i}^q}{q-1}
\label{eq:TsallisDefinition}
\end{equation}

\noindent which, for $ q \rightarrow 1 $ goes to the Shannon
information entropy and for $q=2$ to 1 minus the Onicescu
information energy.

We choose the specific value of the entropy index $q$ for atoms,
using the Tsallis prescription \cite{PrivTsallisPanos}.
Specifically, $T_{q}$ is plotted as a function of $\ln Z$ and the
best value of $q$ is found when the closest trend to a linear
relation of $T_{q}$ with $\ln Z$ is obtained.

\section{Methodology and Numerical Results}
%==========================================
\label{sec:Methodology}

We construct a fractional occupation probability distribution \{$
{p_{i}}\}, i=1,2,\ldots ,\nu $ of electrons in atomic orbitals for
each neutral atom and the corresponding value of $Z$, in a way
similar to our previous work for atomic nuclei (\cite{PaperCP}).
For example, the electron configuration for Ca ($Z=20$) is

\[
    1s^2 \, 2s^2 \, 2p^6 \, 3s^2 \, 3p^6 \, 4s^2
\]

We obtain the corresponding normalized probability distribution
$p_{i}$ by dividing the superscripts (number of electrons
occupying the particular orbitals) by $Z$, i.e. $
p_{1}=p_{1s}=\frac{2}{20} $, $ p_{2}=p_{2s}=\frac{2}{20} $,
$p_{3}=p_{2p}=\frac{6}{20}$, e.t.c. Then, the values \{$p_{i}$\},
summing up to 1, are inserted for the fixed value of $Z=20$ into
previous relations (1) to (10) and the results are shown in Table
1, where $S_{i}=-p_{i}\ln p_{i}$,
$D{i}=(p_{i}-\frac{1}{\nu})^{2}$, $E_{i}=p_{i}^{2}$,
$I_{i}=\frac{(p_{i+1}-p_{i})^{2}}{p_{i}}$ and are summed up at the
last row of the Table. This Table is presented for pedagogical
reasons, so that anybody can understand the simplicity of the
method described here and reproduce our results for all values of
$Z$.
%\ref{tab:pinakas} gives 3 instead of 1.

\begin{figure}[h]
    \centering
    \includegraphics*[scale=0.75]{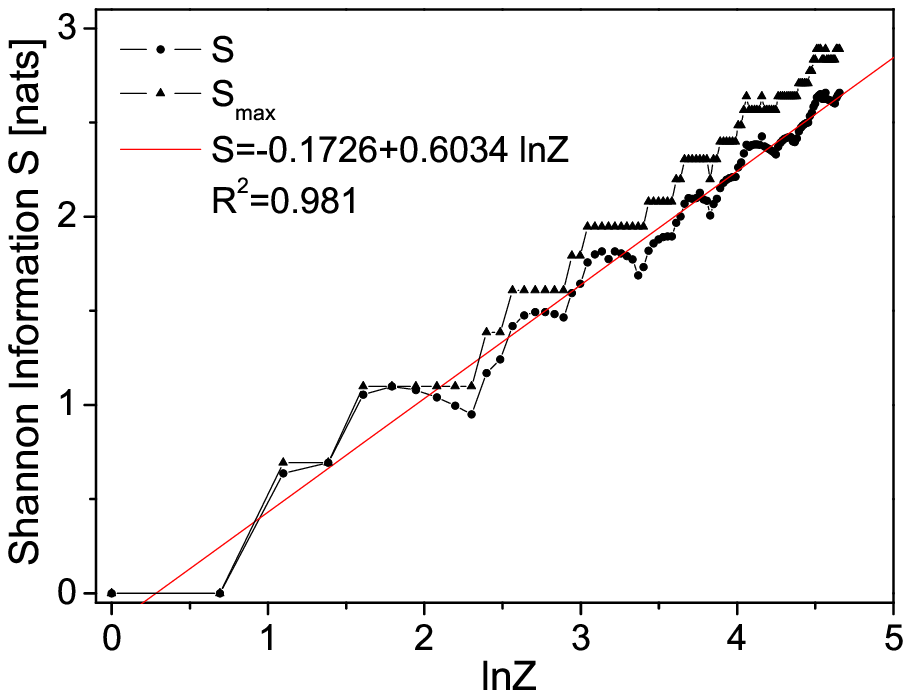}
    \caption[Shannon information entropy]{$S(Z)$, $S_{max}(Z)$ and the fitted expression $S(Z)=a_{S}+b_{S}\,lnZ$}
    \label{fig:graphSh}
\end{figure}

\begin{figure}[h]
    \centering
    \includegraphics*[scale=0.75]{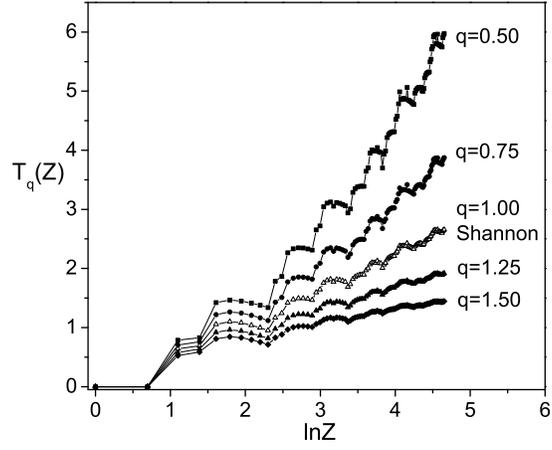}
    \caption[Tsallis information entropy]{Tsallis information entropy $T_{q}(Z)$ for various values of $q$}
    \label{fig:graphTsallis}
\end{figure}

\begin{figure}[h]
\centering
\includegraphics*[scale=0.75]{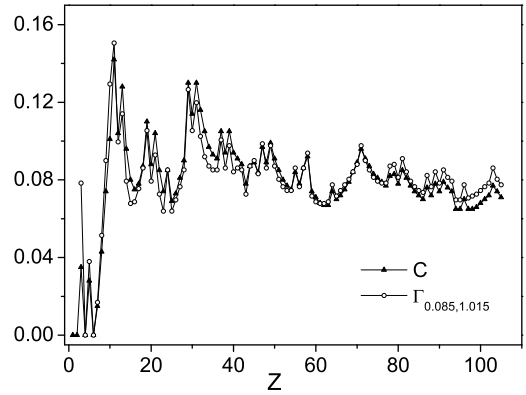}
\caption[PODI]{SDL ($\Gamma_{\alpha,\beta}$) and LMC ($C$) complexities for the calculated $\alpha$ and $\beta$ coefficients (PODI)}
\label{fig:FigabCalc}
\end{figure}

\begin{figure}[h]
    \centering
    \includegraphics*[scale=0.75]{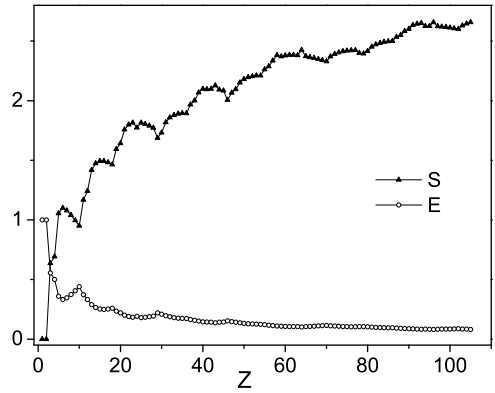}
    \caption[Onicescu]{Onicescu information energy $E(Z)$ and Shannon information $S(Z)$.}
    \label{fig:graphOnSh}
\end{figure}

\begin{figure}[h]
    \centering
    \includegraphics*[scale=0.75]{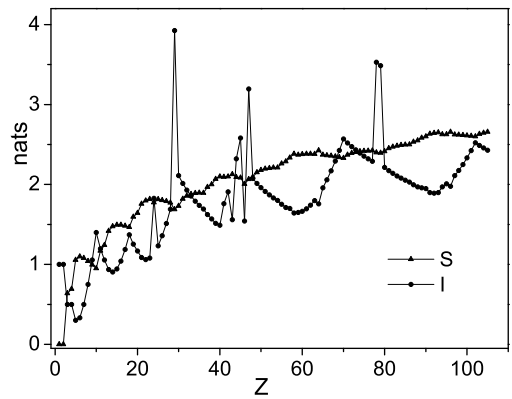}
    \caption[Fisher]{Fisher information $I(Z)$ and Shannon information $S(Z)$}
    \label{fig:graphFiSh}
\end{figure}

\begin{figure}[h]
    \centering
    \includegraphics*[scale=0.75]{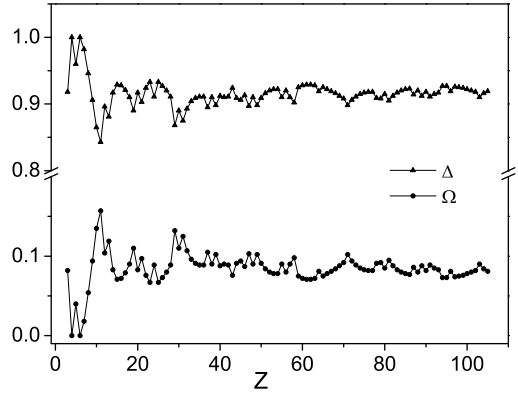}
    \caption[Order and Disorder]{Order $\Delta(Z)$ and Disorder $\Omega(Z)$}
    \label{fig:graphOrderDis}
\end{figure}

\begin{figure}[h]
    \centering
    \includegraphics*[scale=0.75]{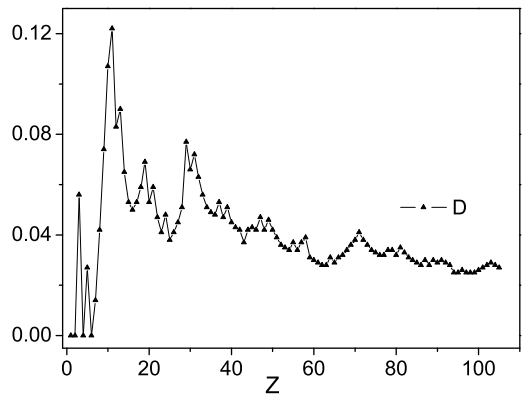}
    \caption[Disequilibrium]{Disequilibrium $D(Z)$}
    \label{fig:graphD}
\end{figure}

Fractional occupation probabilities \{$p_{i}$\} for $Z$=1 to 105,
obtained as described above, are employed as an input into the
formulas of section \ref{sec:Definitions}. Thus, we obtain the
functions $S(Z)$, $S_{max}(Z)$ (Fig.~\ref{fig:graphSh}), $I(Z)$
(Fig.~\ref{fig:graphFiSh}), $\Delta(Z)$, $\Omega(Z)$
(Fig.~\ref{fig:graphOrderDis}), $\Gamma_{\alpha,\beta}(Z)$, $C(Z)$
(Fig.~\ref{fig:FigabCalc}), $D(Z)$ (Fig.~\ref{fig:graphD}), $E(Z)$
(Fig.~\ref{fig:graphOnSh}) and $T_{q}(Z)$
(Fig.~\ref{fig:graphTsallis}). A detailed table with numerical
values for the above functions is available in an online version
of the present
paper\footnote{http://www.autom.teithe.gr/niknik/PaperPCN.pdf}.

\subsection{Shannon Information Entropy}
\label{subsec:Shannon}

\begin{figure}[h]
\centering
\includegraphics*[scale=0.75]{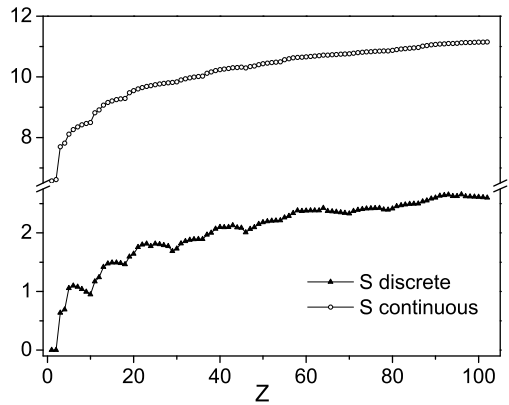}
\caption[S discrete continuous]{$S(Z)$ for discrete (present) and continuous cases (previous work \cite{PaperCMP})} \label{fig:ScSd}
\end{figure}

Employing the maximization of $R^{2}$ (correlation coefficient)
criterion, we calculate the best fit of the form $S(Z)=a_{S} +
b_{S} \ln Z $ to our numerical results for $S(Z)$. The resulting
linear relation is $S_{ln}(Z)=-0.1726+0.6034 \ln Z$, plotted
together with $S(Z)$ and $S_{max}(Z)$ in Fig. \ref{fig:graphSh}.
The two latter curves correspond to our present approach employing
fractional occupation probabilities and are compared in Fig.
\ref{fig:ScSd}, with $S(Z)$ obtained previously \cite{PaperCMP}
with H-F densities (continuous) in position- and momentum- spaces.

\subsection{Tsallis Information Entropy}
\label{subsec:Tsallis}

Special attention is devoted to the calculation of Tsallis entropy
(\ref{eq:TsallisDefinition}) with the entropy index q as a
parameter. The plots of $T_{q}(Z)$ for $q$=0.5, 0.75, 1, 1.25, and
1.5 are shown in Fig. \ref{fig:graphTsallis}.

\begin{figure}[h]
    \centering
    \includegraphics*[scale=0.75]{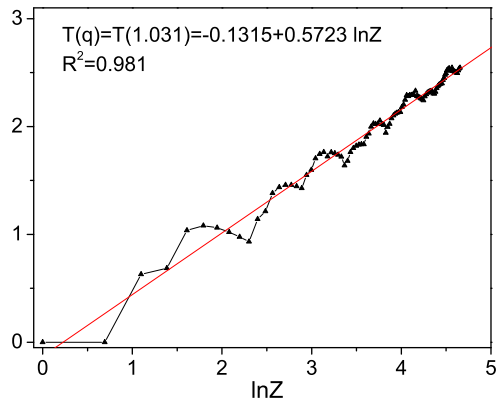}
    \caption[Best fit]{Best fit of $T_{q}(Z)$ with $q=1.031$}
    \label{fig:graphTsallisRegr}
\end{figure}

\begin{figure}[h]
    \centering
    \includegraphics*[scale=0.75]{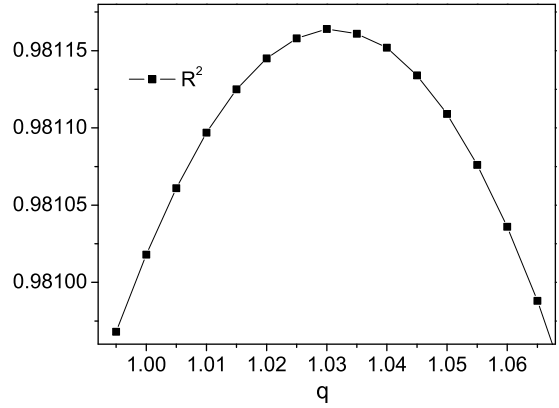}
    \caption[Plot of R2 vs q]{Plot of $R^{2}$ vs $q$, which justifies our choice for $q=1.031$, as optimal}
    \label{fig:graphR2vsQ}
\end{figure}

We assume that $T_{q}(Z)$ is a linear function of $\ln Z$ of the
form $T_{q}(Z) = a_{T} + b_{T} \ln Z$. Thus we follow the Tsallis
prescription \cite{PrivTsallisPanos} described in Section
\ref{sec:Definitions}. We calculate numerically the value of $q$,
which gives the greatest correlation value $R^{2}$, for the linear
fit of $T_{q}(Z)$. The best fit value is $q=1.031$ (Fig.
\ref{fig:graphTsallisRegr} and Fig. \ref{fig:graphR2vsQ}, where
$R^{2}$ is plotted versus q) and the fitted expression is $ T_{ln}
= -0.13147 + 0.57229 \ln Z $. This value of $q$ is close to $q=1$,
corresponding to the Shannon information entropy. We can conclude
that atoms are extensive systems, in the sense that deviation of
$q$ from 1 indicates non-extensivity \cite{Tsallis.paper}. This
"optimal" value of $q$ is obtained easily, employing our simple
method with discrete $\{p_{i}\}$, to be contrasted with the use of
H-F densities, which would involve integrals of the continuous
quantity $p^q(\textbf{r})$.

\subsection{The SDL complexity coefficients $(\alpha,\beta)$ defined as
a Pair of Order-Disorder Indices (PODI)} \label{subsec:SDLcoeff}

As mentioned in Sec. \ref{sec:Definitions}, we may calculate the
particular values of $\alpha$ and $\beta$ by imposing the
condition that the data sets of $\Gamma_{\alpha,\beta}$ and $C$
are approximately the same, in the sense that the sum of squared
errors $\Sigma \epsilon^{2}$ between these two data sets, by
varying the $\alpha$ and $\beta$, coefficients is minimum. The
values calculated are $\alpha = 0.085$ and $\beta = 1.015$ and the
corresponding curves for $\Gamma_{\alpha,\beta}(Z)$ and $C(Z)$ are
plotted together in Fig. \ref{fig:FigabCalc}. The similarity of
the two patterns is obvious. These values can be compared with our
rough guess ($\alpha \simeq 0$, $\beta \simeq 4$) in
\cite{PaperPCMK} which, however, were obtained with continuous H-F
atomic densities, jointly, in position- and momentum-spaces. Our
prescription for the proper values of $(\alpha,\beta)$ enables us
to define them as a Pair of Order-Disorder Indices (PODI), which
quantifies the contribution of order versus disorder to the
complexity of any quantum system \cite{ChatzPanosPODI}.

\begin{figure}[h]
  \begin{center}
    {\label{fig:Gamma1_1}\includegraphics*[scale=0.5]{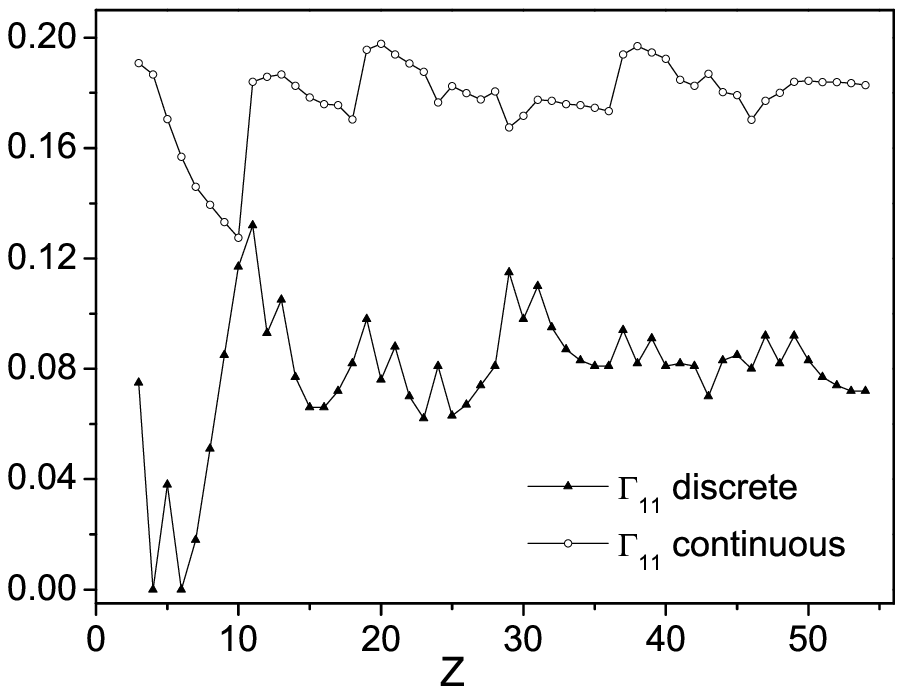}}
    {\label{fig:Gamma1_025}\includegraphics*[scale=0.5]{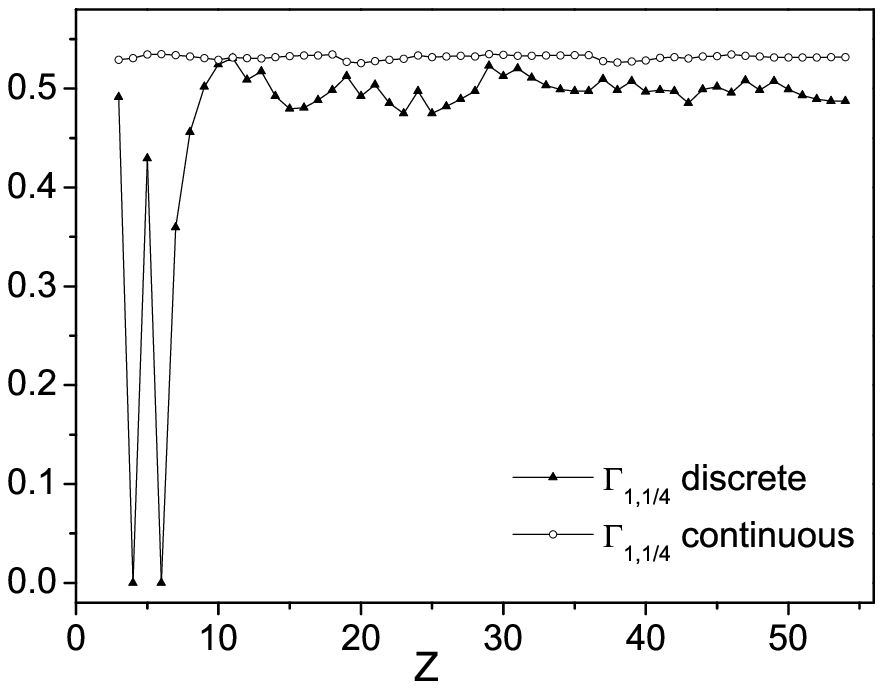}}
    {\label{ContGamma025_0}\includegraphics*[scale=0.5]{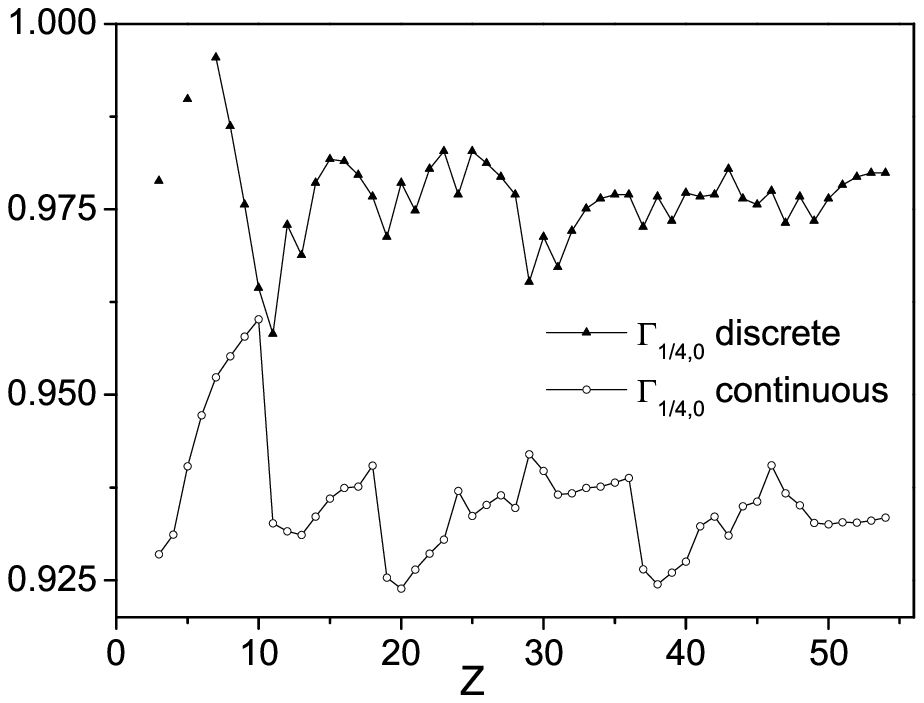}}
    {\label{ContGamma0_4}\includegraphics*[scale=0.5]{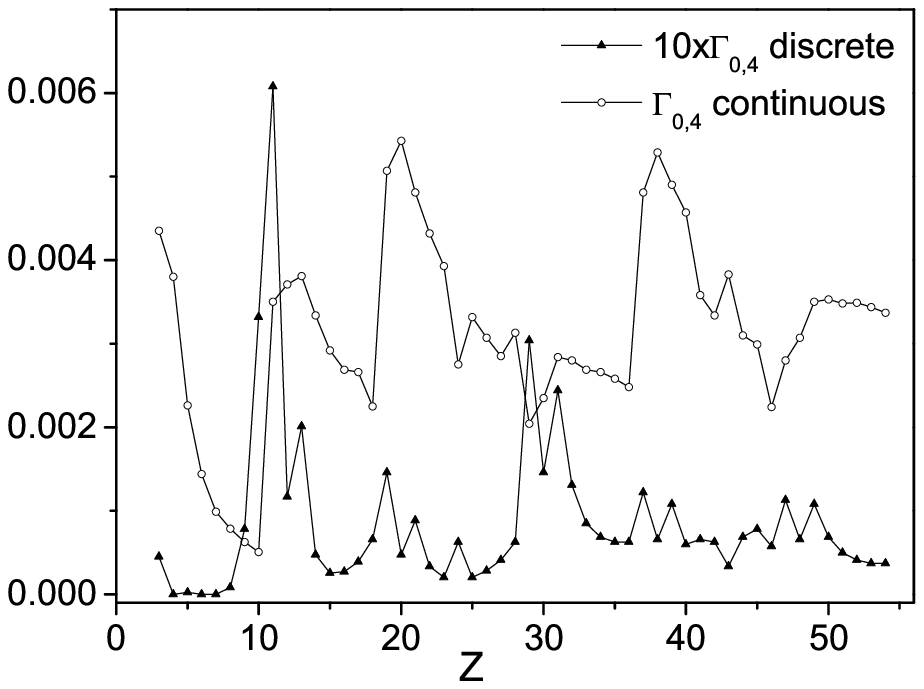}}
  \end{center}
  \caption[SDL 4]{SDL complexity $\Gamma_{\alpha,\beta}(Z)$ for 4 typical pairs of ($\alpha,\beta$)}
  \label{fig:SDL4}
\end{figure}

In Fig. \ref{fig:SDL4} we plot $\Gamma_{\alpha,\beta}(Z)$ for four
typical pairs ($\alpha$,$\beta$), i.e. (1,1), (1, $\frac{1}{4}$),
($\frac{1}{4}$,0) and (0,4) specified in \cite{SDL.paper} and
compare with the corresponding curves of \cite{PaperCMP}. It is
seen that both, continuous and discrete cases, lead on the average
to the same conclusion, that the atom cannot grow in complexity as
$Z$ increases, by observing all the values of $Z$. The same trend
has been observed in \cite{PaperCMP}, in contrast to
\cite{PaperPCMK}, where an increasing trend was obtained. This
difference can be resolved by stating that, if one observes the
closed shells, as in \cite{PaperPCMK}, the trend is increasing,
while by inspecting all atoms, the trend is, on the average, not
increasing, as in \cite{PaperCMP}. In \cite{BorgooPaper} the
authors used improved electron densities in position-space by
introducing relativistic corrections and state that complexity
increases with $Z$ for position-space. An analogous work leading
to the same conclusion is \cite{SanudoPaper}. However, we may note
that such results cannot be considered complete or conclusive,
because a proper treatment of this issue should involve both,
position- and momentum- spaces. We mention, for example, the well
known seminal research of Gadre et al \cite{GadreSearsEtAlPaper},
\cite{SenEdBook} and further more recent work \cite{PaperMP2},
\cite{PaperCMP}, where both spaces are taken into account for $S$.
We stress that the resulting momentum-space integrals should be
treated carefully for numerical accuracy.

The final pair ($\alpha=0.085, \beta=1.015$) characterizes the
strength of order of atoms $\beta$ versus the strength of disorder
$\alpha$ throughout the periodic table and can serve as a Pair of
Order-Disorder Indices (PODI) for any quantum many body-system. We
observe that $\beta \gg \alpha$, which indicates that atoms are
"ordered" systems, or more accurately, they are more "ordered"
than "disordered". This fits well with the fact that we can
visualize the creation of atoms by putting electrons one-by-one in
well defined orbitals. Our final conclusion is: Atoms are
"ordered" systems, which do not grow in complexity with $Z$, at
least in the context of the present work. Perhaps, future research
might check and/or modify our present conclusion, if more accurate
densities are employed and treated properly. The merit of our
definition for the PODI pair ($\alpha,\beta$) is under
investigation, by a comparative study of its application in
several quantum systems \cite{ChatzPanosPODI}.

\section{Discussion}
\label{sec:Discussion}

It is seen in Fig. \ref{fig:ScSd} that all local minima of $S(Z)$
obtained in the present work occur at the same values of $Z$, as
the corresponding discontinuities of the slope of $S(Z)$ found in
\cite{PaperCMP} with the more accurate continuous
Roothaan-Hartree-Fock (RHF) wavefunctions
\cite{RHF_AWF_Values.paper}. This fits well with our expectations
for atoms with closed shells of electrons, which can be considered
as more \emph{compact} than neighboring atoms and are expected to
display smaller values of information content. Specifically, in
the present work, we obtain local maxima for $S(Z)$ for $Z$=6, 15,
23, 25, 35, 40, 43, 58, 62, 64, 77, 93, 96, and 105  and local
minima for $Z$=10, 18, 24, 29, 36, 41, 46, 59, 63, 70, 79, 94, and
102.

\begin{figure}[h]
    \centering
    \includegraphics*[scale=0.75]{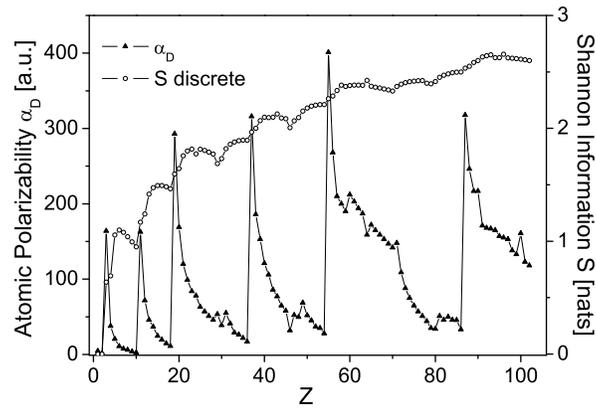}
    \caption[Shannon vs Polarizability]{Shannon information $S(Z)$ and atomic static dipole polarizability $a_{D}$}
    \label{fig:graphShDiPo}
\end{figure}

\begin{figure}[h]
    \centering
    \includegraphics*[scale=0.75]{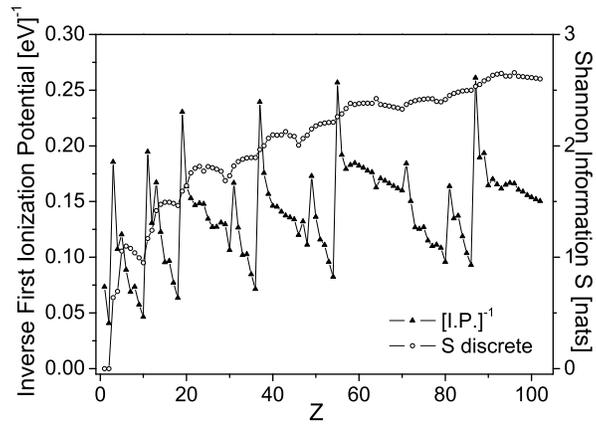}
    \caption[Shannon vs Ion Potential]{Shannon information $S(Z)$ and the inverse first ionization potential $I.P.$}
    \label{fig:graphShIoPo}
\end{figure}

\begin{figure}[h]
    \centering
    \includegraphics*[scale=0.75]{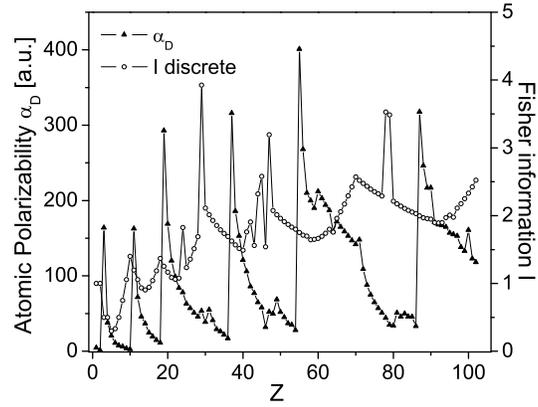}
    \caption[Fisher vs Polarizability]{Fisher information $I(Z)$ and atomic static dipole polarizability $a_{D}$}
    \label{fig:fisherpol}
\end{figure}

\begin{figure}[h]
    \centering
    \includegraphics*[scale=0.75]{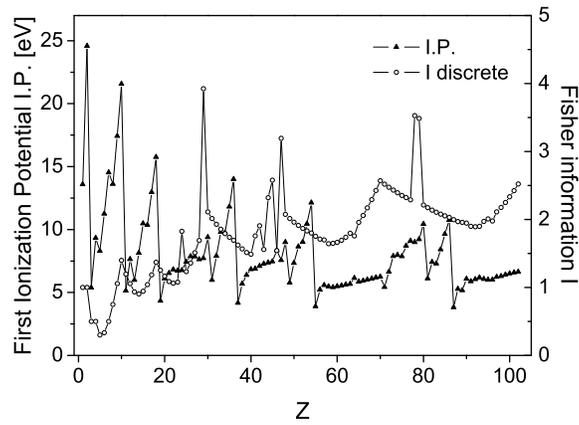}
    \caption[Fisher vs Ion Potential]{Fisher information $I(Z)$ and first ionization potential $I.P.$}
    \label{fig:fisherip}
\end{figure}

We can also observe in Fig. \ref{fig:graphShDiPo} and Fig.
\ref{fig:graphShIoPo} that $S(Z)$ obtained with our simplified
method correlates well with experimental data for atomic dipole
polarizability $\alpha_{D}$ \cite{DipPolBook} and the inverse of
the first ionization potential of atoms $I.P.$ \cite{IonPotSite}.
It is seen that, for closed shells atoms, the local minima of
$\alpha_{D}$ (and, accordingly, the local maxima of $I.P.$)
coincide either with the local minima of $S$ for $Z=10$, $18$ and
$36$ or with discontinuities in the slope of $S$ for $Z=54$ and
$86$. In Fig. \ref{fig:fisherpol} and Fig. \ref{fig:fisherip}, we
also display the correlation of Fisher information $I(Z)$ with
$\alpha_{D}$ and I.P. respectively. In order to appreciate the
significance of those Figures, one should take into account the
reciprocal behavior of $S(Z)$ and $I(Z)$.

In Fig. \ref{fig:graphOnSh} we see the expected reciprocal
behavior of $S(Z)$ and Onicescu information $E(Z)$, while in Fig.
\ref{fig:graphFiSh} a local measure of information, Fisher
information $I(Z)$, is compared with a global one, $S(Z)$. Here,
one verifies that Fisher Information is a more sensitive measure
than $S(Z)$, providing more detailed structure, as a function of
$Z$. It is obvious from our Figures , that $I(Z)$ shows local
maxima in the form of abrupt increases for values of $Z$,
corresponding to just mild discontinuities of the slope of $S(Z)$
(continuous) and local minima of $S(Z)$ (discrete).

Finally, the Disequilibrium $D(Z)$ is shown in Fig.
\ref{fig:graphD}. An additional bonus of that figure is that
$D(Z)$ shows the same pattern (obvious by simple inspection) as
$\Gamma_{\alpha,\beta}(Z)$ and $C(Z)$ of Fig. \ref{fig:FigabCalc}.
Thus, a third measure of complexity emerges.

Our present approximate approach is promising as a first
information-theoretic step to accomplish the transition from one
level of nature to the next, e.g. from atoms to molecules,
indicating an analogous treatment for molecules. Also, our
proposal is pragmatic and contributes towards a unification of
physical systems, from the point of view of self-organization
(organized complexity) using information theory based on a
probabilistic description with minimal computational cost.

The number of particles of atoms, seen as systems of electrons, is
mostly $Z \simeq 10^2$. One might try, for example, to evaluate,
by simplifying, $S$ and $C$ for systems with electrons up to
$Z=10^6$ in ultra-large-scale electronic structure theory
\cite{TakeoHoshiPaper}.

A possible quibble might be the following. Here, we consider
fractional occupation probabilities of electronic configurations
in atomic orbitals. The standard, maybe ideal, method would be,
first, to diagonalize the density matrix
$\rho(\textbf{r},\textbf{r}')$ of electrons, in order to obtain
the corresponding natural occupation numbers and natural orbitals,
and, second construct ${{p_{i}}}, i = 1, 2,\ldots,\nu$ to be
inserted into the formulas. Thus, the effect of electron
correlation could be included in the density matrix. Again, we may
argue that our alternative approach is much more feasible and is,
in this sense, a better candidate for scalability for larger
systems. Here, by "larger" we mean systems with more constituent
particles or entities. It may also serve as a change of our way of
thinking about complexity: one must not hesitate to carry out
calculations for complex systems, but, instead, has to try to find
working effective approximations in order to quantify complexity.

We note that both definitions of complexity,
$\Gamma_{\alpha,\beta}$ (SDL) and $C$ (LMC) are rather not final,
but in view of the inability to give an ideal or single
definition, we might say that both capture basic and desirable
features, which a good measure of complexity should show. This is
usually the case in physics, in order to start thinking about a
new concept. Furthermore, we calculate the optimal pair
($\alpha$,$\beta$) (disorder versus  order parameter), by
requiring that two different and reasonable functions
$\Gamma_{\alpha,\beta}(Z)$ and $C(Z)$ exhibit approximately the
same pattern. The fact that the obtained patterns are extremely
similar, enhances the reliability of our approach.

\section{Conclusions}
\label{sec:Conclusions}

We propose a simplified method to quantify the information content
and complexity in atoms. It is validated by obtaining significant
similarities between several basic features (experimental and
theoretical) in comparison with more sophisticated approaches.
This procedure may be tested for scalability, by applying it to
more complicated systems than the atom, i.e. with more components.
We find the entropic index $q=1.031$, indicating that atoms are
extensive systems. We also present a prescription to find the
proper values of the strength of disorder $\alpha$ and order
$\beta$, defined as a Pair of Order-Disorder Indices (PODI)
characterizing any quantum many-body system \cite{ChatzPanosPODI}.
It is seen that for atoms $\beta \gg \alpha $. Finally, we
conclude that, at least in the context of a non-relativistic
treatment of atoms, taking into account both, position- and
momentum- spaces: "Atoms are "ordered" systems, which do not grow
in complexity as $Z$ increases".

We note that the above conclusion has been tested for neutral
atoms in non-excited states, employing two measures of complexity:
SDL and LMC and a third one, the Disequilibrium $D$, emerging from
this work.

\section*{Acknowledgments} \label{sec:Ackn}
K. Ch. Chatzisavvas is supported by a Post-Doctoral Research
Fellowship of the Hellenic State Institute of Scholarships (IKY).

%%%%%%%%%%%%%%%%%%%%%%%%%%%%%%%%%%%%%%%%%%%%%%%%%%%%%%%%%%%%%%%%%%
%%%%%%%%%%%%%%%%%%%%%%%%%%% References %%%%%%%%%%%%%%%%%%%%%%%%%%%
%%%%%%%%%%%%%%%%%%%%%%%%%%%%%%%%%%%%%%%%%%%%%%%%%%%%%%%%%%%%%%%%%%

%%%%%%%%%%%%%%%%%%%%%%%%%%%%%%%%%%%%%%%%%%%%%%%%%%%%%%%%%%%%%%%%%%
%%%%%%%%%%%%%%%%%%%%%%%%%%%%% Figures %%%%%%%%%%%%%%%%%%%%%%%%%%%%
%%%%%%%%%%%%%%%%%%%%%%%%%%%%%%%%%%%%%%%%%%%%%%%%%%%%%%%%%%%%%%%%%%
\clearpage
\section*{Figures}

%%%%%%%%%%%%%%%%%%%%%%%%%%%%%%%%%%%%%%%%%%%%%%%%%%%%%%%%%%%%%%%%%%
%%%%%%%%%%%%%%%%%%%%%%%%%%%%% Tables %%%%%%%%%%%%%%%%%%%%%%%%%%%%%
%%%%%%%%%%%%%%%%%%%%%%%%%%%%%%%%%%%%%%%%%%%%%%%%%%%%%%%%%%%%%%%%%%
\clearpage
\section*{Tables}
\begin{table}[hbt]
\begin{center}
\begin{tabular}{ccccccccc}
\hline \hline  Element & \multicolumn{3}{c}{Orbital} &
$p_{i}$ & $S_{i}$ & $D_{i}$ & $E_{i}$ & $I_{i}$  \\
\hline  \multicolumn{1}{c}{  } & \multicolumn{3}{c}{$1s^2$} &
0.1000 & 0.2303 & 0.0044 & 0.0100 & 0.0000  \\
 \multicolumn{1}{c}{  } & \multicolumn{3}{c}{$2s^2$} &
0.1000 & 0.2303 & 0.0044 & 0.0100 & 0.4000 \\
 \multicolumn{1}{c}{Ca} & \multicolumn{3}{c}{$2p^6$}&
0.3000 & 0.3612 & 0.0178 & 0.0900 & 0.1333  \\
 \multicolumn{1}{c}{Z=20} &
\multicolumn{3}{c}{$3s^2$}& 0.1000 & 0.2303 & 0.0044 & 0.0100 & 0.4000  \\
 \multicolumn{1}{c}{$\nu$=6} &
\multicolumn{3}{c}{$3p^6$} &
0.3000 & 0.3612 & 0.0178 & 0.0900 & 0.1333  \\
 \multicolumn{1}{c}{  } & \multicolumn{3}{c}{$4s^2$} &
0.1000 & 0.2303 & 0.0044 & 0.0100 & 0.1000  \\
\cline{2-9}
\multicolumn{1}{c}{  } & \multicolumn{3}{c}{Totals} & 1.0000 & 1.6434 & 0.0533 & 0.2200 & 1.1667  \\
\hline \hline
\end{tabular}
\label{tab:pinakas}
\caption[Detailed calculations for Ca]{Results of detailed calculations for Ca}
\end{center}
\end{table}

\end{document}